\documentclass[aps,prl,twocolumn,superscriptaddress,showpacs]{revtex4}

\usepackage{graphicx}% Include figure files
\usepackage{dcolumn}% Align table columns on decimal point
\usepackage{bm}% bold math
\usepackage{color}
\begin{document}

\title{Ultrafast electron-phonon-magnon interactions at noble metal-ferromagnet interfaces}

\author{V. Shalagatskyi}
\affiliation{IMMM CNRS 6283, Universit\'e du Maine, 72085 Le Mans
cedex, France}
\author{O. Kovalenko}
\affiliation{IMMM CNRS 6283, Universit\'e du Maine, 72085 Le Mans
cedex, France} \affiliation{IPCMS CNRS 7504, Universit\'e de
Strasbourg, BP 43, 23 rue du Loess, 67034 Strasbourg Cedex 02,
France}
\author{V. Shumylo}
\affiliation{IMMM CNRS 6283, Universit\'e du Maine, 72085 Le Mans
cedex, France}
\author{A. Alekhin}
\affiliation{IMMM CNRS 6283, Universit\'e du Maine, 72085 Le Mans
cedex, France} \affiliation{Fritz-Haber-Institut der MPG, Phys.
Chemie, Faradayweg 4-6, 14195 Berlin, Germany}
\author{G. Vaudel}
\affiliation{IMMM CNRS 6283, Universit\'e du Maine, 72085 Le Mans
cedex, France}
\author{T. Pezeril}
\affiliation{IMMM CNRS 6283, Universit\'e du Maine, 72085 Le Mans
cedex, France}
\author{V. S. Vlasov}
\affiliation{IMMM CNRS 6283, Universit\'e du Maine, 72085 Le Mans
cedex, France} \affiliation{Syktyvkar State University named after
Pitirim Sorokin, 167001 Syktyvkar, Russia}
\author{A. M. Lomonosov}
\affiliation{LAUM CNRS 6613, Universit\'e du Maine, 72085 Le Mans
cedex, France}
\author{V. E. Gusev}
\affiliation{LAUM CNRS 6613, Universit\'e du Maine, 72085 Le Mans
cedex, France}
\author{D. Makarov}
\affiliation{Helmholtz-Zentrum Dresden-Rossendorf e. V., Institute
of Ion Beam Physics and Materials Research, 01328 Dresden,
Germany}
\author{V. V. Temnov}
\email{vasily.temnov@univ-lemans.fr} \affiliation{IMMM CNRS 6283,
Universit\'e du Maine, 72085 Le Mans cedex, France}
\affiliation{Fritz-Haber-Institut der MPG, Phys. Chemie,
Faradayweg 4-6, 14195 Berlin, Germany}

\date{\today}

\begin{abstract}
Ultrafast optical excitation of gold-cobalt bilayers triggers the
nontrivial interplay between the electronic, acoustic, and
magnetic degrees of freedom. Laser-heated electrons generated at
the gold-air interface diffuse through the layer of gold and
strongly overheat the lattice in cobalt resulting in the emission
of ultrashort acoustic pulses and generation of exchange-coupled
magnons. Time-resolved optical measurements allow for extracting
the thermal boundary (Kapitza) resistances at metal/metal
interfaces and the hot electron diffusion length in ferromagnetic
materials. Both the experimental data and the analytical treatment
of the two-temperature model reveal the role of the Kapitza
resistance in transient lattice overheating.
\end{abstract}

%\pacs{72.55.+s, 75.78.Jp, 43.35.+d, 75.60.Jk, 78.20.hc}
%\keywords{}

\maketitle

Ultrafast heat transport in nanophotonic devices
\cite{CahillJAP2003} plays a crucial role in modern
nanotechnology, in particular in the quest of information
processing at ultrahigh (Tbit/second) rates. For example, thermal
effects determine the fundamental speed limits in nanophotonic
devices based on semiconductor quantum dots when excited by trains
of femtosecond laser pulses with THz repetition rate
\cite{Gomis08PRL101}. In magnetic materials, time-resolved
investigations are largely motivated by the possibility of the
ultrafast manipulation of the magnetic order at the nanoscale
\cite{Beaurepaire96PRL76,vanKampen2002}, recently using the
non-polarized \cite{Eschenlohr2013} and spin-polarized hot
electrons at metal-ferromagnet interfaces
\cite{Melnikov2011PRL107,SchellekensNComm2017,RazdolskiNComm2017,AlekhinPRL2017}.
Besides magnetization dynamics these metal-ferromagnet multilayer
structures were used to manipulate ultrafast plasmonic and
acoustic excitations
\cite{TemnovNaturePhot2012,TemnovNatureComm2013}.

The goal of this Letter is to fill the gap in the understanding of
ultrafast heat transport across metal-ferromagnet interfaces in
multilayer structures. The mechanisms of ultrafast demagnetization
by superdiffusive currents of non-polarized hot electrons in Au/Ni
bilayers \cite{Eschenlohr2013} has been recently questioned based
on the results of transient reflectivity measurements in Pt/Au
bilayers \cite{ChoiPRB2014}. Motivated by earlier works relating
the mechanisms of heat transport by laser-generated hot electrons
and acoustic pulses \cite{Tas94PRB49,Lejman14JOSAB31}, here we
carry out a unifying study on metal-ferromagnet Au/Co bilayers by
performing complimentary measurements of acoustic pulseshapes and
magnetization dynamics. We unambigously demonstrate the phenomenon
of transient overheating of cobalt by hot electrons injected from
gold, which results in generation of (i) ultrashort acoustic
pulses and (ii) coherent exchange-coupled magnons in cobalt.
Analysis and numerical modeling of the thermal transport within
the framework of the two-temperature model reveal the importance
of the interface (Kapitza) thermal boundary resistance
\cite{KapitzaJP1941,CahillJAP2003,SchreierPRB2013,ChoiPRB2014,HahnPRB2015}
in the generation of acoustic pulses and transient overheating.

The inset in Fig.~1(a) illustrates a typical experimental
configuration. Hybrid multilayer structures consist of a 30-nm
hcp-cobalt film grown on a (0001) sapphire substrate and capped
with much thicker (111) gold layer using magnetron
sputtering\cite{TemnovNatureComm2013,TemnovJOPT2016}. Gold
thicknesses were systematically varied between 70 and 270~nm.
Measurements performed in the 'front side pump - back side probe'
geometry are particularly well suited for monitoring the dynamics
of the elementary excitations (electrons, phonons, moagnons)
across heterostructures and buried interfaces.

The gold-air interface is excited by an ultrashort laser pump
pulse (Ti:Sa, 800~nm, 100~fs) and the evolution of surface
reflectivity at the cobalt-sapphire interface is probed by a
time-delayed probe pulse (400~nm, 100~fs). Hot electrons are
generated within the 13-nm-thick optical penetration depth of pump
pulses in gold. The energy transport of these hot electrons is
dominated by diffusion \cite{GusevPRB1998,KaltenbornPRB2012}. The
heat diffusion length in gold during the sub-picosecond
electron-phonon relaxation time
$\zeta_{Au}\sim\sqrt{\kappa/g}\sim 100$~nm \cite{GusevPRB1998} is
determined by the electronic heat conductivity
$\kappa^{(Au)}\simeq 316$~W/mK and the electron-phonon coupling
constant $g^{(Au)}\simeq 2.3\times
10^{16}$~W/m$^{3}$K\cite{Lin08PRB77}. A fraction of hot electrons
that reached the Au/Co interface penetrate into cobalt, where the
heat diffusion length
$\zeta_{Co}=\sqrt{\kappa^{(Co)}/g^{(Co)}}\sim 10$~nm is much
shorter due to smaller heat conductivity $\kappa^{(Co)}\simeq
70$~W/mK and much larger electron-phonon coupling constant
$g^{(Co)}\simeq 60\times 10^{16}$~W/m$^{3}$K
\cite{Bigot05ChemPhys318}. This drastic difference in
electron-phonon coupling leads to the transient overheating of the
cobalt lattice as compared to the adjacent gold. After a few
picoseconds, when thermal equilibrium between the electrons and
the lattice is established, the overheated cobalt cools down by
thermal diffusion into both adjacent layers, i.e. gold and
sapphire.

\onecolumngrid

\begin{figure}[h]
\includegraphics[width= 18cm]{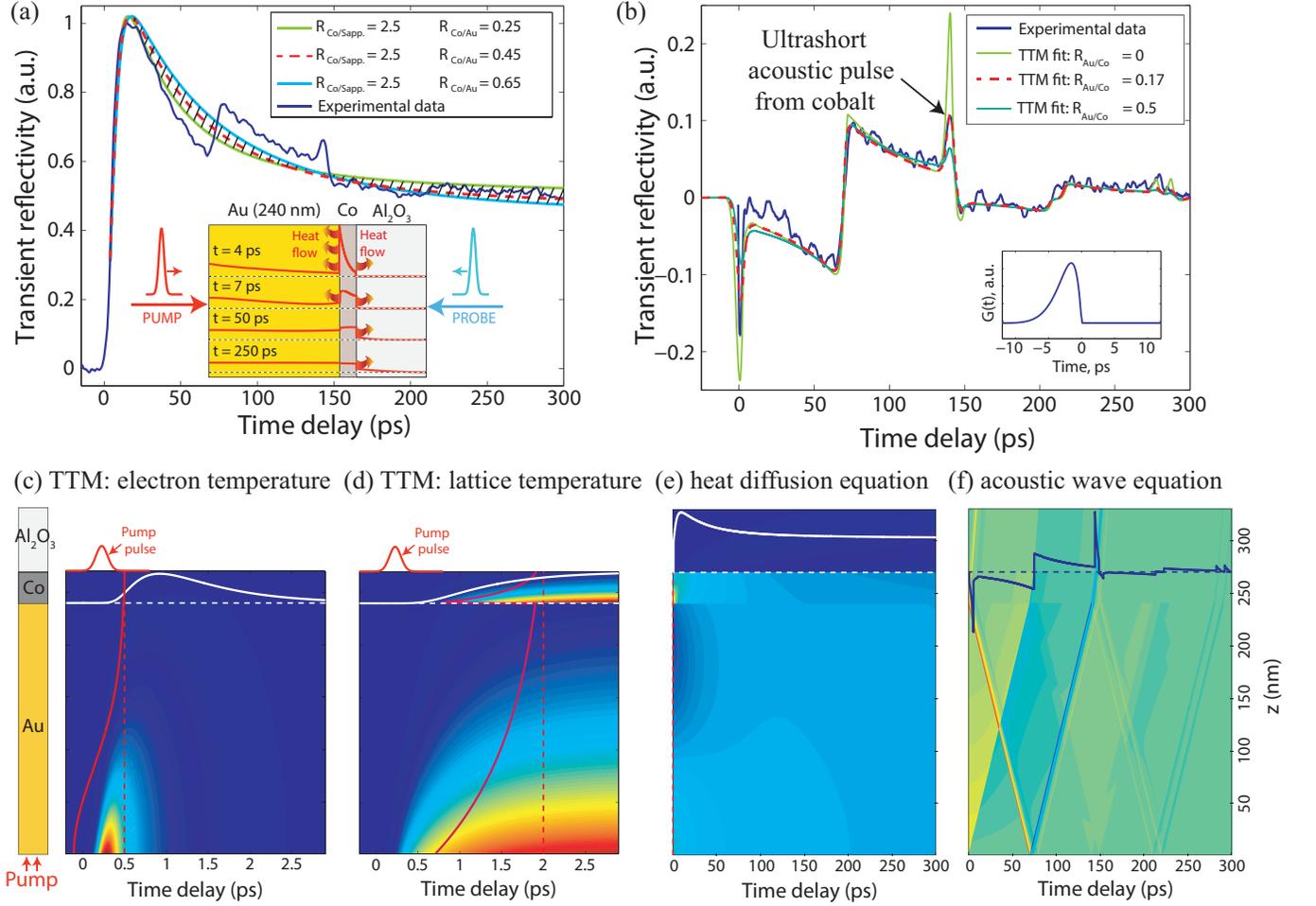}
\caption{(a) Pumping the Au side and probing the Co side of a
Au/Co/sapphire structure, transient reflectivity measurement
indicates strong overheating $\alpha\simeq 15$ of the Co lattice.
Subsequent cooling of cobalt on a picosecond time scale is
dominated by the interfacial thermal (Kapitza) resistance at the
Au/Co interface $R_{Co/Au}<<R_{Co/Sapp.}$; $\alpha$ and
$R_{Co/Au}$ are used as fit parameters. The inset shows solutions
of heat diffusion equation at different delay times. After the
subtraction of the thermal background the residual reflectivity
signal (b) displays the acoustic dynamics with a remarkable
ultrashort acoustic pulse emitted from cobalt. The inset shows the
sensitivity function of the opto-acoustic detection at the
Co/sapphire interface at the probe wavelength. The two-temperature
model (TTM) provides the spatio-temporal evolution of (c) electron
and (d) lattice temperatures on a picosecond time scale. The
transient lattice temperature at $\sim$2~ps after the pump
excitation triggers the dynamics in a (single-temperature) heat
diffusion equation (e) and acoustic wave equation (f) on a longer
time scale.  Solid colored lines in (c-f) display the temporal
(horizontal) and spatial (vertical) profiles (along the dashed
lines) of transient electron and lattice temperatures and acoustic
strain, respectively.}
 \label{fig:ReflectivityFit}
\end{figure}

\twocolumngrid

A representative time trace of the time-resolved reflectivity
measurements is shown in Fig.~1(a). The solution of a heat
diffusion equation generates transient temperature profiles for
different delay times in the inset of Fig.~1(a), where the
direction of heat flow is indicated by the arrows and their number
illustrates the strength of the heat flow. An assumption of a very
unusual initial temperature profile, where the transient cobalt
temperature is at the Au/Co interface is 15$\times$ larger as
compared to the adjacent gold,  was the only feasible possibility
to explain the observed dynamics on a 100~ps time scale in
Fig.~1(a) (see later in the manuscript). The cooling rate of a few
nanometer thin solid layer into adjacent substrates is determined
rather by the interface thermal (Kapitza) boundary
resistances\cite{KapitzaJP1941,CahillJAP2003,HahnPRB2015} than
bulk thermal conductivities. The 100~ps  thermal decay in
Fig.~1(a) can be  modeled assuming fast heat flow from cobalt into
gold and quantified by the Kapitza resistance
$R_{Co/Au}=0.25...0.65$~m$^2$K/GW. A larger value of Kapitza
resistances $R_{Co/Sapp.}\simeq 2.5$~m$^2$K/GW at the
cobalt-sapphire interface was extracted from the measurements on
cobalt thin films (see the Supplemental Material). This value is
in line with the general trend of order-of-magnitude larger
interfacial boundary resistance at the metal-dielectric interfaces
as compared to metal-metal interfaces
\cite{CahillJAP2003,Gundrum05PRB72,ChoiPRB2014}. An important
conclusion of our study is that the cooling rate of a
Au/Co/sapphire structure into the sapphire substrate is determined
by the $R_{Co/Sapp.}<<R_{Au/Sapp.}\simeq 20$~m$^2$K/GW.
Significant difference between $R_{Au/Sapp.}$ and $R_{Co/Sapp.}$
highlights the role of the mismatch in phonon spectra of the metal
and the dielectric substrate, characterized by different Debye
temperatures (see the Supplemental Material) and allows for
engineering of thermal transport at metalA/dielectric interfaces
by sandwiching a thin layer of a phononically matched metal B in a
metalA/metalB/dielectric structure.

After the subtraction of slow thermal dynamics (dashed curve in
Fig.~1(a)), the residual transient reflectivity signal in
Fig.~1(b) explicitly demonstrates the generation of an acoustic
pulse. The most striking experimental observation is the
occurrence of a short spike at 140~ps delay time, which can be
uniquely identified as the acoustic pulse generated at the cobalt
layer and detected after one acoustic roundtrip through the 240~nm
thick gold layer.

In order to provide a qualitative description of the complex
behavior of the transient reflectivity in Fig.~1(a,b) and verify
if the observed thermal and acoustic dynamics is consistent with
the model of the heat transport by laser-induced hot electrons, we
have performed simulations within the framework of the
two-temperature model (TTM).  When applied to metal excited by an
ultrafast laser pulse, the TTM elucidates the dynamics of
electronic and lattice temperatures governed by the following
equations \cite{GusevPRB1998}:
\begin{eqnarray}
\label{TTM} \ C_e(T_e)\frac{\partial T_e}{\partial t} &=& \kappa
(T_e)\frac{\partial^2 T_e}{\partial z^2} - g(T_e - T_l)\nonumber
\\
\ C_l\frac{\partial T_l}{\partial t} &=& g(T_e - T_l)
\end{eqnarray}
where $C_e(T_e)$ is the electronic heat capacity (assumed to
depend linearly on $T_e$ in our range of excitations
\cite{Gundrum05PRB72}), $\kappa(T_e)$ is the temperature-dependent
electronic thermal conductivity, $C_l$ is the lattice heat
capacity.

The diffusive heat transport by hot electrons from gold into
cobalt is strongly affected by the phenomenological (electronic)
Kapitza resistance at the Co/Au interface
\begin{eqnarray}
\label{KapitzaResistance}
R_{Au/Co}=-\frac{T^{(Au)}_e-T^{(Co)}_e}{\kappa(\partial{T_e}/\partial{z})}\,,
\end{eqnarray}
which results in a discontinuity of the electronic temperature at
the Au/Co interface and reduces the heat flow from Au to Co.
Taking into account that heat transport in metals is dominated by
electrons, we neglect possible contributions of phononic Kapitza
resistance. The analytical solution of the TTM equations (see the
Supplemental Material) leads to the following expression for the
overheating of cobalt
\begin{eqnarray}
\label{TemperatureJump} \alpha=\frac{\Delta T^{(Co)}_l}{\Delta
T^{(Au)}_l} =
\frac{g^{(Co)}C^{(Au)}_l}{g^{(Au)}C^{(Co)}_l}\frac{1}{[1+R_{Au/Co}\sqrt{\kappa^{(Co)}g^{(Co)}}]}\,,
\end{eqnarray}
once electron-phonon relaxation process is completed. Given a
large ratio $g^{(Co)}/g^{(Au)}\sim 25$ this estimate provides a
strong overheating $\alpha\simeq 4...19$ for the range of Kapitza
resistances between 0.5 and 0~m$^2$K/GW, respectively.

Attempting to fit the experimental data in Fig.~1(a,b), we found
that it is impossible to fit both the acoustic and 100~ps cooling
dynamics with the same fit parameter $R_{Au/Co}=R_{Au/Co}$. Trying
to fit the acoustic amplitude and cooling dynamics separately, the
Kapitza resistance  for the acoustics $R_{Au/Co}=0.17$~m$^2$K/GW
was found to be smaller than $R_{Co/Au}=0.45$~m$^2$K/GW obtained
using the single-temperature model to reproduce cooling of the
cobalt film. Taking into account that the amplitude of a
picosecond acoustic pulse after one roundtrip through the gold
layer becomes smaller due to the acoustic attenuation would
require $R_{Au/Co}$ to be further reduced.

Moreover, a smaller value of $R_{Au/Co}<0.2$~m$^2$K/GW needs to be
put into Eq.~(3) in order to justify the overheating factor
$\alpha\simeq 15$ required to obtain a correct value of transient
reflectivity at 300~ps delay time in Fig.~1(a). Whereas the
fitting of the acoustic pulse amplitude with a smaller Kapitza
resistance may be regarded with certain skepticism, a large and
constant overheating factor $\alpha\simeq 15$, unambiguously
confirmed for structures with different gold thickness (see the
Supplemental Material), represents a strong argument for the
asymmetry in heat transport.

This asymmetry, manifesting itself as $R_{Au/Co}\neq R_{Co/Au}$,
may have several reasons. Possible phononic contributions to the
Kapitza resistance at metal-metal interfaces can be safely
disregarded as the heat transport in metals is dominated by
electrons  \cite{SchreierPRB2013}. The previously reported
decrease of the thermal boundary resistance at the metal-metal
interface at elevated temperatures \cite{Gundrum05PRB72} is not
evident as the overheating factor $\alpha$ was found to not depend
on the thickness of the gold layer.

We conclude that the TTM modeling, when applied to fit the thermal
and acoustic dynamics, explains the phenomenon of the strong
overheating of cobalt quantified by a large overheating factor
$\alpha\simeq 15$. The observed asymmetry in the Kapitza
resistance at metal-metal interfaces remains subject of further
investigations, perhaps within the framework of the microscopic
description of thermal rectification \cite{HahnPRB2015}.
Reflectivity measurements on similar Au/Fe/MgO-samples display a
weaker acoustic peak from iron, which is explained by a larger
value of the Kapitza resistance $R_{Fe/Au}\simeq 0.9$~m$^2$K/GW
and a smaller overheating factor $\alpha\simeq 5$ (see the
Supplemental Material). A comparative analysis of
$R_{metal/metal}$ within the framework of the diffusive mismatch
model \cite{Gundrum05PRB72} would be required to understand the
role of electronic densities of states in both metals.

\begin{figure}[h]
\includegraphics[width= 8cm]{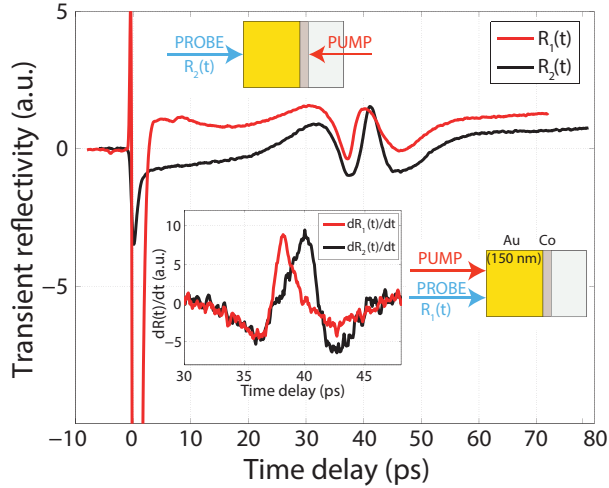}
\caption{Reflectivity measurements at the gold-air interface
evidence the generation of ultrashort acoustic pulses resembling
the heat penetration profile in cobalt (see the inset and
Ref.~\cite{Manke13APL103} for details).}
\label{fig:ReflectivityFit}
\end{figure}

Coming back to the analysis of the Co/Au structures we note that
the emission of ultrashort acoustic pulses from the overheated
cobalt film can be also observed in the transient reflectivity
signal probing the Au/air interface, and pumping the multilayer
structure either through the substrate or from the Au side (see
Fig. 2). We have shown recently that the time-derivative of simple
reflectivity measurements matches the acoustic pulses shape if it
is small compared to the 3.8~ps acoustic travel time through the
skin depth in gold at 400~nm probe wavelength
\cite{Manke13APL103}. The time derivatives of the transient
reflectivity in the inset of Fig.~2 compare the acoustic pulse
shapes generated by the direct optical absorption in cobalt
(pumping the Co/sapphire interface) and by hot electrons through
gold (pumping Au/air interface). Both pulse shapes are consistent
with the exponential heat penetration. A 3~ps pulse generated by
the direct optical excitation of cobalt corresponds to the heat
penetration depth of 20~nm \cite{TemnovNatureComm2013}, which is a
convolution between the optical penetration and the hot electron
heat diffusion depth. A $\tau_{Co}$=2~ps short acoustic pulse upon
exciting the Au/air interface directly measures the hot electron
diffusion depth $\zeta_{Co}=\tau_{Co}c_s=12$~nm ($c_s=6$~nm/ps is
the sound velocity in hcp-cobalt).

The experimental observation of 2~ps short acoustic pulses
generated in cobalt by thermo-elastic mechanism unambiguously
demonstrate the spatial localization of transient lattice heating,
which is known to excite some non-trivial magnetization dynamics
\cite{vanKampen2002,RazdolskiNComm2017}. Figure 3(a) shows
temporal evolution of the Kerr rotation at the Co-sapphire
interface measured under the external magnetic field of 350~mT
tilted by 45 degrees from the surface normal. Analogous to the
experiments with direct optical excitation of ferromagnetic films
\cite{vanKampen2002}, here we observe magnetization oscillations
at different frequencies. Whereas in our case the magnetization
oscillations are triggered solely by hot electrons, the physical
interpretation is nearly identical to that by van Kampen and
co-workers ~\cite{vanKampen2002}. The magnetization precession in
Fig. 3(a) is quantitatively reproduced by the superposition of two
damped oscillations: a low-frequency ferromagnetic resonance (FMR,
$n=0$) and high-frequency first-order magnon mode ($n=1$). Two
experimental observations corroborate our conclusion that a
high-frequency oscillation can be interpreted as a standing mode
of exchange coupled magnons in cobalt. First, the high frequency
oscillation is shifted by $\pi$ with respect to FMR-oscillation.
Taking into account the sinusoidal spatial distribution of
magnetization in a first-order magnon mode
$M(z)=\cos(\pi\frac{z}{L})$ (see the inset in Fig.~3(a)) we
conclude that the Kerr rotation measurement within the optical
skin depth at the cobalt-sapphire interface should result in a
$\pi$-phase shift. Second, the frequency $\omega_1$ of the
high-frequency oscillation agrees well with the estimations for
the exchange-coupled magnon in cobalt:
$\hbar\omega_1\simeq\hbar\omega_{\rm FMR}+D(\pi/L)^2$ with
$D\simeq550$~meV$~\AA^2$ taken from Ref. \cite{Liu1996}. In some
samples we were able to observe a short-living second-order
($n=2$) magnon mode at a frequency of
$\hbar\omega_2\simeq\hbar\omega_{\rm FMR}+D(2\pi/L)^2$ (see the
Supplemental Material for exact expressions).

\begin{figure}[h]
\includegraphics[width= 6cm]{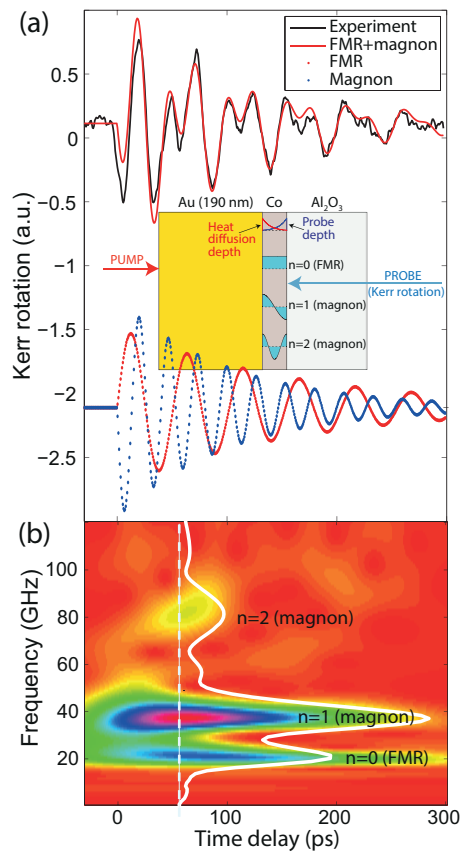}
\caption{(a) Spatially inhomogeneous heating of cobalt triggers a
complex magnetization dynamics consisting of the FMR-precession
(21~GHz,~$n=0$) and exchange-coupled magnon (37~GHz,~$n=1$). (b)
Continuous wavelet transform of Kerr rotation signal evidences the
excitation of a weak short-living second order ($n=2$) magnon at
80~GHz.} \label{fig:ReflectivityFit}
\end{figure}

Excitation of n=2 magnons is evidenced in a wavelet transform of
the time-dependent Kerr rotation signal (Fig.~3(b)), which,
analogous to a windowed Fourier transform, provides information
about the temporal evolution of the signal spectrum
\cite{Tomassini01AO40}. The comparison of excitation amplitudes of
the first ($n=1$) and second-order ($n=2$) magnons $\sim
((L/\zeta_{Co})^2+(\pi n)^2)^{-1}$ as well as their detection
efficiencies $\sim ((L/\delta_{skin})^2+(\pi n)^2)^{-1}$ give a
ratio of 15$\%$, in agreement with our experimental data. The
dynamics of exchange-coupled magnons can also be affected by
magnetoelastic interactions with acoustic pulses, a new phenomenon
to be reported elsewhere.

To summarize, we have explored non-equilibrium pathways of energy
transport in gold-cobalt bilayers mediated by hot electron
diffusion through a macroscopically thick gold during the time of
electron-phonon thermalization. A strong and spatially
inhomogeneous transient overheating of cobalt results in the
emission of ultrashort acoustic pulses and excitation of coherent
exchange-coupled magnons (with n=1,2). Analytical modeling within
the framework of the two-temperature model quantifies the role of
the Kapitza resistance at the Au/Co-interface, which determines
the degree of overheating, amplitude of emitted acoustic pulses
and the initial cooling dynamics back into the gold layer on a
picosecond time scale. The new ultrafast pathway of remote energy
deposition  strongly localized at a specific area in the bulk of
excited structure is expected to play a dramatic role in
laser-induced nanofabrication through femtosecond laser-induced
phase transitions \cite{SokolowskiTintenPRL1998} occurring at high
irradiation powers.

The authors are indebted to D. Mounier for assistance with
nanosecond reflectivity measurements, to D. Diesing for providing
a Au/Fe/MgO sample, to V. Besse, P. Ruello, A. Melnikov and I.
Razdolski for discussions and to R. Tobey for useful suggestions
and his help in editing the manuscript. Funding through {\it
Nouvelle \'{e}quipe, nouvelle th\'{e}matique} "Ultrafast acoustics
in hybrid magnetic nanostructures", {\it Strat\'{e}gie
internationale NNN-Telecom and the Acoustic HUB de la R\'{e}gion
Pays de La Loire}, {\it Alexander von Humboldt Stiftung}, the
European Research Council (FP7/2007-2013) / ERC grant agreement
no. 306277 and PRC CNRS-RFBR "Acousto-magneto-plasmonics" (grant
number 1757 150001) is greatfully acknowledged.

%\bibliographystyle{unsrt}
%\bibliography{BibliographyKapitza}

\end{document}